\documentstyle[twoside,  epsf, hyperref]{article}

\input ibvs3.sty

\begin{document}
\IBVShead{6243}{13 July 2018}

\IBVStitletl{New transit timing observations for}{GJ~436~b, HAT-P-3~b, HAT-P-19~b, WASP-3~b, and XO-2~b}

\IBVSauth{MACIEJEWSKI,~G.$^1$; STANGRET,~M.$^1$; OHLERT,~J.$^{2,3}$; BASARAN,~\c{C}.S.$^4$; MACIEJCZAK,~J.$^5$; PUCIATA-MROCZYNSKA,~M.$^5$; BOULANGER,~E.$^5$}

\IBVSinst{Centre for Astronomy, Faculty of Physics, Astronomy and Informatics, 
             Nicolaus Copernicus University, Grudziadzka 5, 87-100 Toru\'n, Poland, e-mail: gmac@umk.pl}
\IBVSinst{Michael Adrian Observatorium, Astronomie Stiftung Trebur, 65468 Trebur, Germany}
\IBVSinst{University of Applied Sciences, Technische Hochschule Mittelhessen, 61169 Friedberg, Germany}
\IBVSinst{Astronomy and Space Sciences Department, Istanbul University, 34116 Fatih Istanbul, Turkey}
\IBVSinst{Zesp\'o\l{} Szk\'o\l{} Uniwersytetu Miko\l{}aja Kopernika, Gimnazjum i Liceum Akademickie, Szosa Che\l{}mi\'nska 83, 87-100 Toru\'n, Poland}

\SIMBADobj{GJ436b}
\SIMBADobj{HAT-P-3b}
\SIMBADobj{HAT-P-19b}
\SIMBADobj{WASP-3b}
\SIMBADobj{XO-2b}
\IBVSkey{photometry}
\IBVSabs{We present new transit observations acquired between 2014 and 2018 for the hot exoplanets GJ~436~b, HAT-P-3~b, HAT-P-19~b, WASP-3~b, and XO-2~b. New mid-transit times extend the timespan covered by observations of these exoplanets and allow us to refine their transit ephemerides. All new transits are consistent with linear ephemerides.}

\begintext

Precise transit timing for an exoplanet may lead to discovering deviations from a linear ephemeris that can be interpreted as a departure from a simple Keplerian model of the planetary orbital motion. Those so called transit timing variations (TTVs) could be induced by unseen planetary companions, such as in the Kepler-19 system (Ballard et al.\ 2011, Malavolta et al.\ 2017), or by nearby low-mass planets in compact planetary systems, such as WASP-47 (Becker et al.\ 2015). Transit timing is also a great tool for studying star-planet interactions (e.g.\ Birkby et al.\ 2014, Ragozzine \& Wolf 2009). For the exoplanet WASP-12~b, a decrease of the orbital period, that can be interpreted as the result of orbital decay due to tidal dissipation inside the star, has been detected (Maciejewski et al.\ 2016). In this research note, we present new transit observations for hot exoplanets in the systems GJ~436 (Butler et al.\ 2004, Gillon et al.\ 2007), HAT-P-3 (Torres et al.\ 2007), HAT-P-19 (Hartman et al.\ 2011), WASP-3 (Pollacco et al.\ 2008), and XO-2 (Burke et al.\ 2007). The photometric time series were used to determine mid-transit times a number of epochs after previous observations available in the literature, and to refine transit ephemerides. We find no deviations from the Keplerian solutions for any  investigated exoplanets.  

The bulk of observations was acquired with the 0.6 m Cassegrain telescope at the Centre for Astronomy of the Nicolaus Copernicus University (Toru\'n, Poland). An SBIG STL-1001 CCD camera was used as detector. The instrumental setup offered a field of view (FoV) of $11\farcm8  \times 11\farcm8$. In order to increase the signal-to-noise ratio for transit timing purposes, observations were acquired ether without any filter (clear mode) or through a blue blocking ($\lambda < 500$ nm) long-pass filter (LP500). The maximum of the spectral response in the LP500 filter was found to be close to the middle of the $R$ band, and for white light the maximum was found to fall between the $V$ and $R$ bands. The transit of XO-2~b on 2014 Mar 20 was observed with the 1.2 m Trebur telescope at the Michael Adrian Observatory (Trebur, Germany). The instrument was equipped with an SBIG STL-6303 CCD camera and provided a $10\farcm0  \times 6\farcm7$ FoV. For that run, photometric measurements were acquired alternately in a Bessel $R$ filter and in white light. To suppress flat-field errors, telescopes were guided manually with an accuracy of a few arc seconds. The timestamps were synchronised to UTC with at least sub-second accuracy via Network Time Protocol. Basic information on the observations are listed in Table~1. 

\vskip 0.5cm
\centerline{Table 1. List of observed transits.}
\vskip 2mm
{\small
\begin{center}
\begin{tabular}{llccccccc}
\hline
Date  &  Telescope & Filter & $t_{\rm{exp}}$ (s) & $N_{\rm{exp}}$ & $N_{\rm{fin}}$ & $pnr$  & Epoch  &  $T_{\rm{mid}}$ (BJD$_{\rm{TDB}}$) \\
          &                   &          &                            &                          &             &    &             &  $+2450000$  \\
\hline
\multicolumn{9}{c}{GJ 436 b} \\
2017 Mar 26 & 0.6 m Toru\'n & LP500 & $15$ & $525$ & $177$ & $1.14$ & $1259$ & $7839.47013^{+0.00052}_{-0.00051}$ \\
2018 Feb 22 & 0.6 m Toru\'n & LP500 & $15$ & $570$ & $189$ & $1.12$ & $1385$ & $8172.60146^{+0.00067}_{-0.00064}$ \\
\hline
\multicolumn{9}{c}{HAT-P-3~b} \\
2017 Mar 28 & 0.6 m Toru\'n & LP500 & $25$ & $506$ & $257$ & $1.01$ & $1249$ & $7840.53170^{+0.00024}_{-0.00022}$ \\
2018 May 07 & 0.6 m Toru\'n & LP500 & $25$ & $598$ & $301$ & $1.10$ & $1389$ & $8246.49585^{+0.00027}_{-0.00028}$ \\
\hline
\multicolumn{9}{c}{HAT-P-19~b} \\
2015 Oct 04 & 0.6 m Toru\'n & clear & $40$ & $354$ & $271$ & $1.84$ & $551$ & $7300.37489^{+0.00042}_{-0.00038}$ \\
2015 Oct 08 & 0.6 m Toru\'n & clear & $50$ & $343$ & $315$ & $1.83$ & $552$ & $7304.38284^{+0.00039}_{-0.00039}$ \\
\hline
\multicolumn{9}{c}{WASP-3~b} \\
2017 Apr 30 & 0.6 m Toru\'n & LP500 & $15$ & $550$ & $190$ & $1.39$ & $2020$ & $7874.4583^{+0.0011}_{-0.0011}$ \\
2018 Apr 15 & 0.6 m Toru\'n & LP500 & $15$ & $637$ & $211$ & $1.85$ & $2209$ & $8223.50975^{+0.00072}_{-0.00078}$ \\
\hline
\multicolumn{9}{c}{XO-2~b} \\
2014 Mar 20 & 1.2 m Trebur & R & $60$ & $146$ & $146$ & $1.34$ & $990$ & $6737.45198^{+0.00038}_{-0.00041}$ \\
                     &                 & clear & $20$ & $146$ & $146$ & $1.85$ &            &  \\
2018 Jan 07 & 0.6 m Toru\'n & LP500 & $25$ & $641$ & $326$ & $1.03$ & $1396$ & $7799.49032^{+0.00032}_{-0.00032}$ \\
2018 Apr 06 & 0.6 m Toru\'n & LP500 & $15$ & $930$ & $318$ & $1.57$ & $1555$ & $8215.41123^{+0.00044}_{-0.00043}$ \\
\hline
\multicolumn{9}{l}{Dates are given in UTC for mid-transit times. $t_{\rm{exp}}$ is the exposure time used. $N_{\rm{exp}}$ is the number of } \\
\multicolumn{9}{l}{scientific exposures recorded. $N_{\rm{fin}}$ is the number of data points in the final light curve after resampling. }\\
\multicolumn{9}{l}{$pnr$ is the photometric scatter in parts per thousand of the normalised flux per minute of observation. }\\
\multicolumn{9}{l}{Epoch is the transit number from the initial time $T_{0}$. $T_{\rm{mid}}$ is the best-fitting mid-transit time.}\\
\end{tabular}
\end{center}
}
\vskip 0.5cm

The observations were subjected to a standard reduction procedure which included dark correction and flat-fielding with sky flats. The magnitudes were obtained with the AstroImageJ package (Collins et al.\ 2017) employing the differential aperture photometry method. Both the aperture size and the set of comparison stars were optimised for individual light curves to achieve the smallest photometric scatter for the target star. Simultaneous detrending against the airmass, position on the matrix, time, and seeing was used if justified. The light curves were normalised to unity outside transits and the timestamps were converted to barycentric Julian dates in barycentric dynamical time ($\rm{BJD_{TDB}}$). The photometric noise rate ($pnr$), defined as the rms per minute of exposure (Fulton et al.\ 2011), was calculated to quantify the quality of each light curve. The final light curves were resampled into 1 minute intervals.  

The Transit Analysis Package (TAP, Gazak et al.\ 2012) was used to derive mid-transit times. The software employs the approach of Mandel \& Agol (2002) to generate transit models and the wavelet-based technique of Carter and Winn (2009) to account for the time-correlated noise.  For the individual systems, their transit parameters such as the inclination and semi-major axis in stellar radii were taken from the literature and allowed to vary under Gaussian penalties determined by parameters' uncertainties. Transit depths, coded by the ratio of planetary and stellar radii, were kept free for the individual light curves in order to account for imperfect de-trending and possible third-light contamination. Tables of Claret \& Bloemen (2011) were explored with the {\tt EXOFAST} applet\footnote{http://astroutils.astronomy.ohio-state.edu/exofast/limbdark.shtml} (Eastman et al.\ 2013) to estimate the limb darkening (LD) coefficients of the quadratic law for the LP500 and $R$ bands, as well as the white light. Stellar parameters were taken from von Braun et al.\ (2012) for GJ 436 and from Torres et al.\ (2012) for HAT-P-3, HAT-P-19, WASP-3, and XO-2. To account for possible inaccuracies in predictions of the LD law (e.g.\ M{\"u}ller et al.\ 2013), the LD coefficients were allowed to vary around the theoretical predictions under the Gaussian penalties equal to 0.1. Since multi-parameter de-trending is not implemented in TAP, we applied a simplified approach in which the intercept and slope of the out-of-transit brightness were allowed to float to account for any remaining trends in the total error budget. The fitting procedure was based on 10 Markov chain Monte Carlo walks with $10^6$ steps each. Median and the 15.9 and 85.1 percentile values of marginalised posteriori probability distributions were taken as the best-fitting values and their 1-$\sigma$ uncertainties. No correlations between the determined mid-transit times, $T_{\rm{mid}}$, and other fitted parameters were found. The results are given in Table~1, and the light curves\footnote{The photometric time series are available online in Tables~3--7} with the best-fitting models are plotted in Fig.~1.

\IBVSfig{15cm}{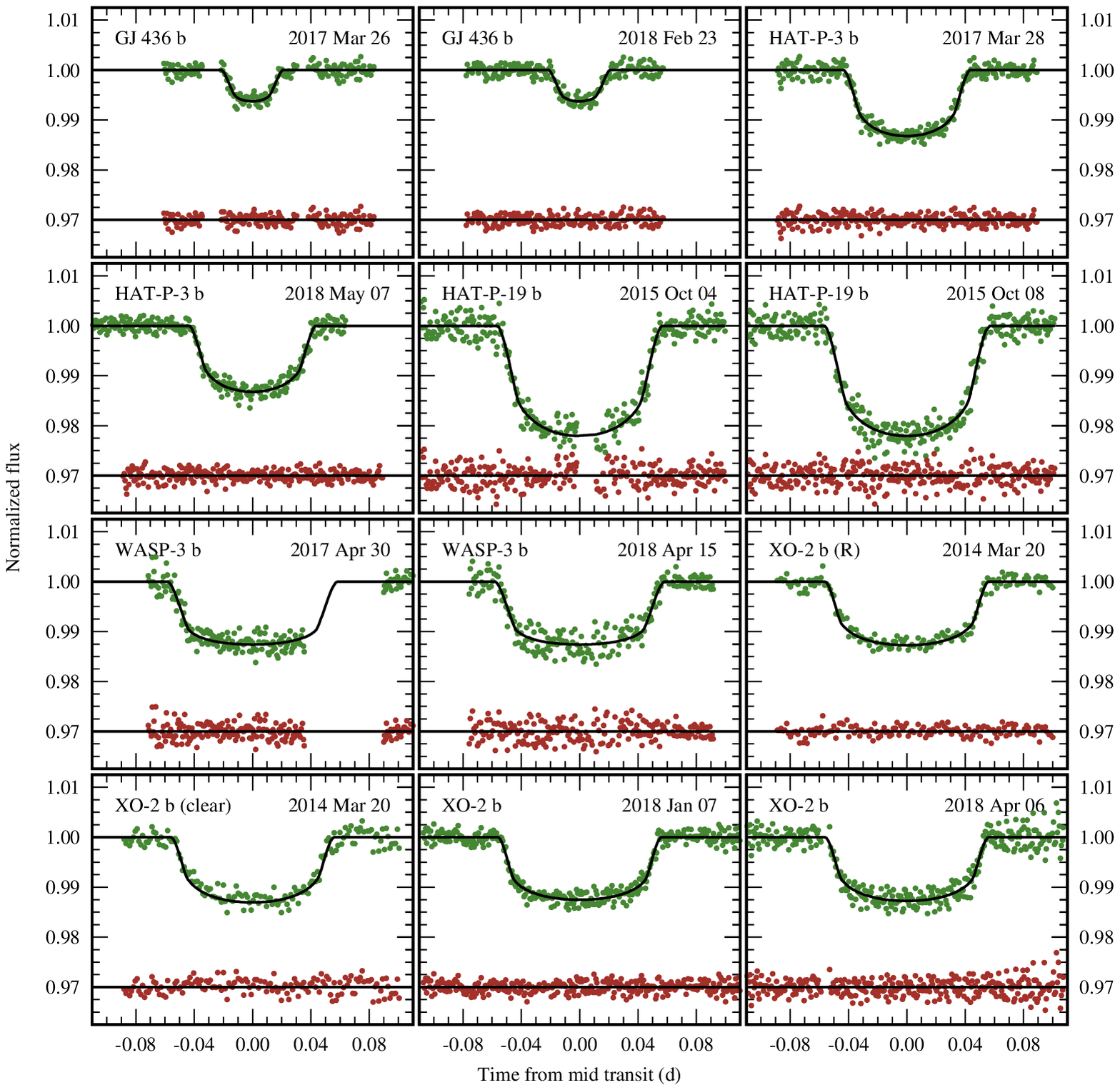}{New transit light curves for GJ 436 b, HAT-P-3~b, HAT-P-19~b, WASP-3~b, and XO-2~b. The continuous lines show the best-fitting transit models. The residuals are plotted below.}

The new mid-transit times were combined with those available in the literature to refine the transit ephemerides in a form 
\begin{equation}
T_{\rm{mid}} = T_0 + P \times E,  \\
\end{equation} 
where $E$ is the transit number starting from the initial epoch $T_0$, which is usually adopted from the discovery paper, and $P$ is the orbital period. The results for the individual exoplanets, together with the goodness of the fit represented by the reduced chi square $\chi^{2}_{\rm{red}}$, are given in Table~2. The timing residuals from the linear ephemerides are plotted in Fig.~2. The new transits are consistent with the refined linear ephemerides for all investigated exoplanets.

\IBVSfig{19cm}{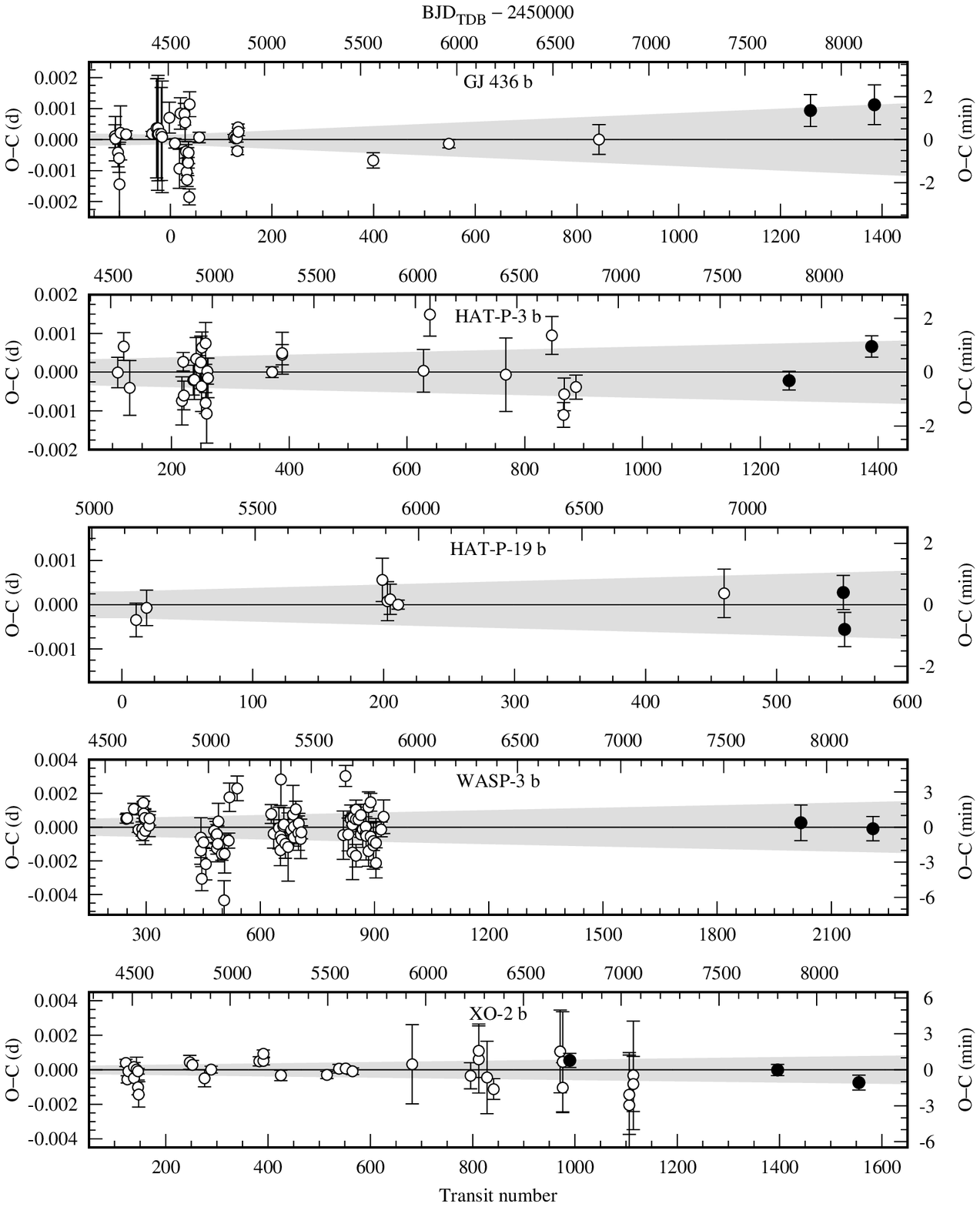}{Transit timing residuals from the linear ephemerides for GJ 436 b, HAT-P-3~b, HAT-P-19~b, WASP-3~b, and XO-2~b. Open circles show literature data, and the filled dots place mid-transit times reported in this research note. The propagation of the ephemerides' uncertainties at the 95.5\% confidence level are marked by grey areas.}

\vskip 0.5cm
\centerline{Table 2. Refined transit ephemerides.}
\vskip 2mm
{\small
\begin{center}
\begin{tabular}{lccc}
\hline
Planet  & $T_0$ $(\rm{BJD_{TDB}})$ & $P$ $(\rm{d})$ & $\chi^{2}_{\rm{red}}$ \\
\hline
GJ 436 b & $2454510.801640 \pm 0.000076$ & $2.64389797 \pm 0.00000040$ & 5.6\\
HAT-P-3 b & $2454218.75960 \pm 0.00016$ & $2.89973764 \pm 0.00000026$ & 2.2\\
HAT-P-19 b & $2455091.53501 \pm 0.00015$ & $4.00878332 \pm 0.00000059$ & 0.73\\
WASP-3 b & $2454143.85112 \pm 0.00022$ & $1.84683510 \pm 0.00000032$ & 3.2\\
XO-2 b & $2454147.75066 \pm 0.00012$ & $2.61585937 \pm 0.00000024$ & 2.0\\ 
\hline
\end{tabular}
\end{center}
}
\vskip 0.5cm

{\it Acknowledgements:} We are grateful to the anonymous referee for comments that helped to clarify some steps of the presented analysis. GM and MS acknowledge the financial support from the National Science Centre, Poland through grant no. 2016/23/B/ST9/00579.

\references

Ballard, S., Fabrycky, D., Fressin, F., et al., 2011, {\it ApJ}, {\bf 743}, 200 \DOI{10.1088/0004-637X/743/2/200}

Becker, J.C., Vanderburg, A., Adams, F.A., et al., 2015, {\it ApJ}, {\bf 812}, L18 \DOI{10.1088/2041-8205/812/2/L18}

Birkby, J.L., Cappetta, M., Cruz, P., et al., 2014, {\it MNRAS},  {\bf 440}, 1470 \DOI{10.1093/mnras/stu343}

Burke, Ch.J., McCullough, P.R., Valenti, J.A., et al., 2007, {\it ApJ}, {\bf 671}, 2115 \DOI{10.1086/523087}

Butler, R.P., Vogt, S.S., Marcy, G.W., et al., 2004, {\it ApJ}, {\bf 617}, 580 \DOI{10.1086/425173}

Carter, J.A., \& Winn, J.N., 2009, {\it ApJ}, {\bf 704}, 51 \DOI{10.1088/0004-637X/704/1/51}

Claret, A., \& Bloemen, S., 2011, {\it A\&A}, {\bf 529}, A75 \DOI{10.1051/0004-6361/201116451}

Collins, K.A., Kielkopf, J.F., Stassun, K.G., et al., 2017, {\it AJ}, {\bf 153}, 77 \DOI{10.3847/1538-3881/153/2/77}

Eastman, J., Gaudi, B. S., \& Agol, E.,  2013,  {\it PASP}, {\bf 125}, 83 \DOI{10.1086/669497}

Fulton, B.J., Shporer, A., Winn, J.N., et al., 2011, {\it AJ}, {\bf 142}, 84 \DOI{10.1088/0004-6256/142/3/84}

Gazak, J.Z., Johnson, J.A., Tonry, J., et al., 2012, {\it Advances in Astronomy}, {\bf 2012}, 697967 \DOI{10.1155/2012/697967}

Hartman, J.D., Bakos, G.\'A., Sato, B., et al., 2011, {\it ApJ}, {\bf 726}, 52 \DOI{10.1088/0004-637X/726/1/52}

Gillon, M., Pont, F., Demory, B.-O., et al., 2007, {\it A\&A}, {\bf 472}, L13 \DOI{10.1051/0004-6361:20077799}

Maciejewski, G., Dimitrov, D., Fern\'andez, M., et al., 2016,  {\it A\&A}, {\bf 588}, L6 \DOI{10.1051/0004-6361/201628312}

Malavolta, L., Borsato, L., Granata, V., et al., 2017, {\it AJ}, {\bf 153}, 224 \DOI{10.3847/1538-3881/aa6897}

Mandel, K., \& Agol, E., 2002, {\it ApJ}, {\bf 580}, L171 \DOI{10.1086/345520} 

M{\"u}ller, H.M., Huber, K.F., Czesla, S., et al., 2013, {\it A\&A}, {\bf 560}, A112 \DOI{10.1051/0004-6361/201322079} 

Pollacco, D., Skillen, I., Collier Cameron, A., et al., 2008, {\it MNRAS}, {\bf 385}, 1576 \DOI{10.1111/j.1365-2966.2008.12939.x}

Ragozzine, D., Wolf, A.S., 2009, {\it ApJ}, {\bf 698}, 1778 \DOI{10.1088/0004-637X/698/2/1778}

Torres, G., Bakos, G.\'A., Kovacs, G., et al., 2007, {\it ApJ}, {\bf 666}, L121 \DOI{10.1086/521792}

Torres, G., Fischer, D.A., Sozzetti, A., et al., 2012, {\it ApJ}, {\bf 757}, 161 \DOI{10.1088/0004-637X/757/2/161}

von Braun, K., Boyajian, T.S., Kane, S.R., et al., 2012, {\it ApJ}, {\bf 753}, 171 \DOI{10.1088/0004-637X/753/2/171}

\endreferences

\end{document}